\documentclass{fac}
\usepackage{graphicx}
\usepackage{amssymb}
\usepackage{amsfonts}
\usepackage{amsmath}
\usepackage{dsfont}
\usepackage{MnSymbol}
\usepackage{mathtools}

\makeatother

\ifprodtf \else \usepackage{latexsym}\fi

\title[An Algebraic Approach for Approximity]
      {An Algebraic Approach for Approximity}

\author[Yong Wang]
    {Yong Wang\\
     School of Computer Science and Technology,\\
     Beijing University of Technology, Beijing, China\\
     }

\correspond{Yong Wang, Pingleyuan 100, Chaoyang District, Beijing, China.
            e-mail: wangy@bjut.edu.cn}

\pubyear{2011}
\pagerange{\pageref{firstpage}--\pageref{lastpage}}

\begin{document}
\label{firstpage}

\makecorrespond

\maketitle

\begin{abstract}
Comparison to traditionally accurate computing, approximate computing focuses on the rapidity of the satisfactory solution, but not the unnecessary accuracy of the solution. Approximate bisimularity is the approximate one corresponding to traditionally accurate bisimilarity. Based on the work of distances between basic processes, we propose an algebraic approach for distances between processes to support a whole process calculus CCS, which contains prefix, sum, composition, restriction, relabeling and recursion.
\end{abstract}


\section{Introduction}\label{Introduction}

Big data is a new trend in computer science, because there has been more and more huge amount of data produced by the various sensors, the Internet and the IoT. For traditionally accurate computing to process these data needs more and more computation power, and also more and more processing time, approximate computing will become a key technology in big data area.

Comparison to traditionally accurate computing, approximate computing focuses on the rapidity of the satisfactory solution, but not the unnecessary accuracy of the solution. Approximate bisimularity is the approximate one corresponding to traditionally accurate bisimilarity. Therefore, that caused to process calculus based approximate semantics \cite{AB} \cite{TOPO} with comparison to traditional CCS \cite{CCS} \cite{CCS2}.

Similar to approximate bisimilarity, distances between processes \cite{SD} \cite{DPS} \cite{DPA} were researched recently. Especially, in \cite{DPA}, an pure algebraic approach for distances between processes was proposed, which could be compared to algebraic laws for processes \cite{HML}.

But, only basic processes (freely generated by $\Sigma_1(\textbf{0},Act,+)$-algebra), i.e., the process terms only contain prefix operator and sum operator. A whole process calculus in \cite{CCS} \cite{CCS2} is not supported in \cite{DPA}. Based on the work in \cite{DPA}, we propose an algebraic approach for distances between processes to support a whole process calculus in \cite{CCS} \cite{CCS2}, which contains prefix, sum, composition, restriction, relabeling and recursion.

This paper are organized as follows. In section 2, we introduce the preliminaries, including the process calculus used in this paper, the approximate bisimulation semantics, and the work of distances between basic processes in \cite{DPA}. In section 3, we discuss distances between processes with silent step $\tau$. We discuss distances between processes with composition operator $\mid$ in section 4. In section 5, distances between processes with restriction and relabeling are discussed. Then in section 6, distances between processes with recursion are introduced. And finally, we conclude this paper in section 7.

\section{Preliminaries}\label{Pre}

In the following, the variables $x,x',y,y',z,z'$ range over the collection of process terms, $p,q$ are processes, the variables $\upsilon,\omega$ range over the set $Act$ of atomic actions, $\alpha,\beta\in A$, $\overline{Act}$ is the set of co-names of actions, $\overline{\alpha},\overline{\beta}\in \overline{Act}$, $t,t'$ are closed items, $\tau$ is the special constant silent step, $\overline{d}(\alpha,\beta)$ is the distance between $\alpha$ and $\beta$.

\subsection{Process Calculus}

\textbf{Definition 1 (Signature)}. A signature $\Sigma$ consists of a finite set of function symbols (or operators) $f,g,\cdots$, where each function symbol $f$ has an arity $ar(f)$, being its number of arguments. A function symbol $a,b,c,\cdots$ of arity \emph{zero} is called a constant, a function symbol of arity one is called unary, and a function symbol of arity two is called binary.

\textbf{Definition 2 (Term)}. Let $\Sigma$ be a signature. The set $\mathbb{T}(\Sigma)$ of (open) terms $s,t,u,\cdots$ over $\Sigma$ is defined as the least set satisfying: (1)each variable is in $\mathbb{T}(\Sigma)$; (2) if $f\in \Sigma$ and $t_1,\cdots,t_{ar(f)}\in\mathbb{T}(\Sigma)$, then $f(t_1,\cdots,t_{ar(f)}\in\mathbb{T}(\Sigma))$. A term is closed if it does not contain free variables. The set of closed terms is denoted by $\mathcal{T}(\Sigma)$.

We have considered as processes the terms generated by the free $\Sigma(\textbf{0},Act\cup\overline{Act}\cup\{\tau\},+,\mid, \backslash,f,\overset{\text{def}}{=})$-algebra, which correspond to the  CCS processes defined in \cite{CCS}\cite{CCS2} and \cite{AB}\cite{TOPO}.

\textbf{Definition 3 (Operational semantics of a process calculus)}. Given a set of actions $\alpha\in Act$, the set of processes is that defined by the BNF-grammar: $p ::= \textbf{0} | \alpha p | p + q | p \mid q | p \backslash L | f(p) | A \overset{\text{def}}{=} P$. The operational semantics is defined by:

$$\textbf{Act}\quad\quad\frac{}{\alpha.p\xrightarrow{\alpha}p}$$

$$\textbf{Sum}\quad\quad\frac{p\xrightarrow{\alpha}p'}{p + q\xrightarrow{\alpha}p'}$$

$$\textbf{Com}_1\quad\quad\frac{p\xrightarrow{\alpha}p'}{p\mid q\xrightarrow{\alpha}p'\mid q}$$

$$\textbf{Com}_2\quad\quad\frac{q\xrightarrow{\alpha}q'}{p\mid q\xrightarrow{\alpha}p\mid q'}$$

$$\textbf{Com}_3\quad\quad\frac{p\xrightarrow{l}p'\quad q\xrightarrow{\overline{l}}q'}{p\mid q\xrightarrow{\tau}p'\mid q'}$$

$$\textbf{Com}_3^\theta\quad\quad\frac{p\xrightarrow{l_1}p'\quad q\xrightarrow{\overline{l_2}}q'}{p\mid q\xrightarrow{\tau}p'\mid q'}\quad\quad(\overline{d}(l_1,l_2)<\theta)$$

$$\textbf{Res}\quad\quad\frac{p\xrightarrow{\alpha}p'}{p\backslash L\xrightarrow{\alpha} p'\backslash L}\quad\quad(\alpha,\overline{\alpha}\notin L)$$

$$\textbf{Rel}\quad\quad\frac{p\xrightarrow{\alpha}p'}{p[f]\xrightarrow{f(\alpha)}p'[f]}$$

$$\textbf{Con}\quad\quad\frac{p\xrightarrow{\alpha}p'}{A\xrightarrow{\alpha}p'}\quad\quad A \overset{\text{def}}{=} p$$

\subsection{Approximate Bisimulation}

The following concepts come from Ying's works on approximate Bisimulation \cite{AB} \cite{TOPO}.

\textbf{Definition 4 ($\lambda$-bisimulation)}. Let $\sigma=(S,T,\{\xrightarrow{t}:t\in T\})$ be a transition system and $\lambda\in [0,\infty)$. $R$ is a $\lambda$-bisimulation, if and only if for any $(s_1, s_2)\in R$, and for each $\lambda<\theta$ and for each $t\in T$,
\begin{enumerate}
  \item if $s_1\xrightarrow{t}s_1'$ then for some $u\in T$, $\exists s_2'\in S$, $s_2\xrightarrow{u}s_2'$, $\overline{d}(t,u)<\theta$ and $(s_1',s_2')\in R$; and
  \item if $s_2\xrightarrow{t}s_2'$ then for some $u\in T$, $\exists s_1'\in S$, $s_1\xrightarrow{u}s_1'$, $\overline{d}(t,u)<\theta$ and $(s_1',s_2')\in R$.
\end{enumerate}

\textbf{Definition 5 ($\lambda$-round, non-expansive)}. Let $(X,\overline{d})$ be a metric space, $Y\subseteq X$ and $\lambda\geq 0$, and let $f$ be a mapping from $X$
into itself. If for any $x,y\in X$, $x\in Y$ and $\overline{d}(x,y)\leq \lambda$ implies $y\in Y$, then $Y$ is said to be $\lambda$-round; if for some $\mu>\lambda$, $Y$ is $\mu$-round, then $Y$ is said to be strongly $\lambda$-round, and if for any $x,y\in X$, $\overline{d}(f(x),f(y))\leq \overline{d}(x,y)$, then $\overline{d}$ is said to be non-expansive.

Let agent schemas $\Omega$ be those expressions in which action variables at different places must be different such that they can be put in with arbitrary actions to construct an agent expression, and $Av(E)$ is the set of action places (occurring syntactically) in $E$.

\textbf{Definition 6 (Weakly $\lambda$-defended)}. Let $E\in\Omega$. If an agent variable $X$ fits the following conditions and $\lambda\geq 0$:
\begin{enumerate}
  \item $X$ does not occur in any subagent of $E$ of the form $E_1\mid E_2$;
  \item if $X$ occurs in a subagent $E'\backslash L$ of $E$; then $L$ is $\lambda$-round; and
  \item if $X$ occurs in a subagent $E'[f]$ of $E$; then $f$ is non-expansive,
\end{enumerate}
then $X$ is said to be $\lambda$-defended. If we eliminate condition 1, then $X$ is weakly $\lambda$-defended.

\subsection{Distances Between Processes}

In \cite{DPS} and \cite{DPA}, the distance relation $=_d$ between $\Sigma_1(\textbf{0},Act,+)$-algebra processes is defined as follows.

\textbf{Definition 7 (Distance relation of $\Sigma_1$-algebra processes)}. Given a semantics $\mathcal{L}$, defined by a pre-order $\subseteq_{\mathcal{L}}$, coarser than bisimulation, we say that a process $q$ is at distance at most $n$ of being better than some other $p$, w.r.t. the semantics $\mathcal{L}$ and the distance between actions $\overline{d}$, and then we write $d^{\mathcal{L}}_{\overline{d}}\leq n\geq 0$, if we can infer $p\sqsubseteq^{\mathcal{L}}_{n}$, by applying the following rules:

1.$\frac{p\subseteq_{\mathcal{L}}q}{p\sqsubseteq^{\mathcal{L}}_{n}q}$.

2.$\frac{p\sqsubseteq^{\mathcal{L}}_{n}q}{\alpha p\sqsubseteq^{\mathcal{L}}_{n+\overline{d}(\beta,\alpha)}\beta q}$.

3.$\frac{p\sqsubseteq^{\mathcal{L}}_{n}p'}{p+q\sqsubseteq^{\mathcal{L}}_{n}p'+q}$.

4.$\frac{p\sqsubseteq^{\mathcal{L}}_{n}q\quad q\sqsubseteq^{\mathcal{L}}_{n'}r}{p\sqsubseteq^{\mathcal{L}}_{n+n'}r}$.

We simply write $=_d$ for the obtained collection of distance relations, that in this case are all symmetric.

\textbf{Definition 8 (Algebraic Definition of the bisimulation distance of $\Sigma_1$-algebra)}. Two processes $p,q\in\Sigma_1$, we can say that $p$ is at most $d\geq 0$ far away of being bisimilar to $q$, if and only if $p\equiv_d q$ can be derived using the following set of rules:

  1. $p \equiv_d p$ for all $d\geq 0$ and for all $p$.

  2. $p \equiv_d p'\Rightarrow p' \equiv_d p$ for all $d\geq 0$ and for all $p,p'$.

  3. $p \equiv_{d_1} p'$ and $p'\equiv_{d_2} p'' \Rightarrow p \equiv_{d_1+d_2} p''$ for all $d_1, d_2\geq 0$ and for all $p,p',p''$.

  4. (i)$p \equiv_{d_1} p'$ and $q\equiv_{d_2} q' \Rightarrow p + q \equiv_{d_1+d_2} p' + q'$ for all $d_1, d_2\geq 0$ and for all $p,p',q,q'$.

      (ii)$p \equiv_d q\Rightarrow \alpha p \equiv_d \alpha q$ for all $d\geq 0$ and for all $p,q$, and $\alpha\in Act$.

  5. $p \equiv_d p'\Rightarrow p\rho \equiv_d p'\rho$ for every substitution $\rho$.

  6. (i)$\alpha x\equiv_{\overline{d}(\alpha,\beta)}\beta x$ for all $\alpha,\beta\in Act$.

      (ii)$x+y\equiv_0 y+x$.

      (iii)$x+x\equiv_0 x$.

      (iv)$(x+y)+z\equiv_0 x + (y+z)$.

      (v)$z+\textbf{0}\equiv_0 z$.

Then in \cite{DPA}, the following conclusion is proven.

\textbf{Theorem 1}. For all $p,q\in \Sigma_1$, $p\equiv_d q \Leftrightarrow p =_d q$.

\section{Distances between Processes with the Silent Step $\tau$}

In this section, we consider $\Sigma_2 (\textbf{0},Act\cup\overline{Act}\cup\{\tau\},+)$-algebra. Because in $\Sigma_2$, the processes are with the co-names $\overline{\alpha}$ of an action $\alpha$, and the silent step $\tau$, we define the following distances between actions.

$\overline{d}(\tau,\tau)=0$.

$\overline{d}(\alpha,\beta)=\overline{d}(\overline{\alpha},\overline{\beta})$.

$\overline{d}(\tau,\beta)=\overline{d}(\alpha,\tau)=\overline{d}(\tau,\overline{\beta})=\overline{d}(\overline{\alpha},\tau)=\overline{d}(\alpha,\overline{\beta})=\overline{d}(\overline{\alpha},\beta)=\infty$.

Because of the existence of $\tau$, the distance relation in $\Sigma_2$ is based on weak approximate bisimulation \cite{TOPO}, which is coarse than weak bisimulation.

\textbf{Definition 9 (Distance relation of $\Sigma_2$-algebra processes)}. Given a semantics $\mathcal{L}$, defined by a pre-order $\subseteq_{\mathcal{L}}$, coarser than weak bisimulation, we say that a process $q$ is at distance at most $n$ of being better than some other $p$, w.r.t. the semantics $\mathcal{L}$ and the distance between actions $\overline{d}$, and then we write $d^{\mathcal{L}}_{\overline{d}}\leq n\geq 0$, if we can infer $p\sqsubseteq^{\mathcal{L}}_{n}$, by applying the following rules:

1.$\frac{p\subseteq_{\mathcal{L}}q}{p\sqsubseteq^{\mathcal{L}}_{n}q}$.

2.$\frac{p\sqsubseteq^{\mathcal{L}}_{n}q}{\alpha p\sqsubseteq^{\mathcal{L}}_{n+\overline{d}(\beta,\alpha)}\beta q}$.

3.$\frac{p\sqsubseteq^{\mathcal{L}}_{n}p'}{p+q\sqsubseteq^{\mathcal{L}}_{n}p'+q}$.

4.$\frac{p\sqsubseteq^{\mathcal{L}}_{n}q\quad q\sqsubseteq^{\mathcal{L}}_{n'}r}{p\sqsubseteq^{\mathcal{L}}_{n+n'}r}$.

We simply write $=_d$ for the obtained collection of distance relations, that in this case are all symmetric.

Next, we define the $\equiv_d$ relation as follows.

\textbf{Definition 10 (Algebraic Definition of the bisimulation distance of $\Sigma_2$-algebra)}. We just add the following rules as 7 to Definition 8:

7. (i)$x+\tau x\equiv_0\tau x$.

(ii)$\alpha(x+\tau y)\equiv_0 \alpha(x+y)+ \alpha y$.

Then we can get the following theorem.

\textbf{Theorem 2}. For all $p,q\in \Sigma_2$, $p\equiv_d q \Leftrightarrow p =_d q$.

\begin{proof}
The proof needs to use induction over the depth of derivations. Comparison to the proof of Theorem 1, we only consider the new cases.

($\Leftarrow$).

The new cases are as follows.

1. $\frac{p\approx q}{p=_d q}$. If $p\approx q$, then by use of the last four axioms in Def.10.6 (same as Def.8.6) and the two axioms in Def.10.7, we have $p\equiv_0 q$, by Def.10.1 we have $q\equiv_d q$, finally by Def.10.3, we get $p\equiv_d q$.

2. $\frac{p=_d q}{\alpha p=_{d+\overline{d}(\beta,\alpha)} \beta q}$. For the cases of $\alpha, \beta\in Act\cup\overline{Act}\cup \{\tau\}$, the conclusion still stands.

($\Rightarrow$).

1. $p\equiv_d p$. We have $p\approx p$, then we get $p=_d p$.

2. The two axioms in Def.10.7, by applying the i.h. and Def.9.1 and Def.9.2.
\end{proof}

\section{Distances between Processes with the Composition Operator $\mid$}

In this section, let us consider $\Sigma_3 (\textbf{0},Act\cup\overline{Act}\cup\{\tau\},+,\mid)$.

\textbf{Definition 11 (Distance relation of $\Sigma_3$-algebra processes)}. We just add the following rules to Definition 9 as rule 5.

5.$\frac{p\sqsubseteq^{\mathcal{L}}_{n}p'}{p\mid q\sqsubseteq^{\mathcal{L}}_{n}p'\mid q}$ for $n<\theta$.

\textbf{Definition 12 (Algebraic Definition of the bisimulation distance of $\Sigma_3$-algebra)}. We just add the following rules as 7 and 8 to Definition 8:

7.(i)$p\equiv_{d_1} p', q\equiv_{d_2} q' \Rightarrow p\mid q\equiv_{d_1+d_2} p'\mid q'$ for $d_1,d_2<\theta$.

(ii)If $u$ is $\sum \mu_i x_i$ and $v$ is $\sum \nu_j y_j$, then $u\mid v\equiv_d\sum \mu_i(x_i\mid v) + \sum \nu_j(u\mid y_j) + \sum_{\overline{d}(\mu_i,\nu_j)<\theta} \tau(x_i\mid y_j)$ (where $d=\overline{d}(\mu_i,\nu_j),\overline{d}(\mu_i,\nu_j)<\theta)$.

8.(i)$x+\tau x\equiv_0\tau x$.

(ii)$\alpha(x+\tau y)\equiv_0 \alpha(x+y)+ \alpha y$.

(iii)$\alpha\tau y\equiv_0 \alpha y$.

(iv)If $u$ is $\sum \mu_i x_i$ and $v$ is $\sum \nu_j y_j$, then $u\mid v\equiv_0\sum \mu_i(x_i\mid v) + \sum \nu_j(u\mid y_j) + \sum_{\mu_i=\overline{\nu_j}} \tau(x_i\mid y_j)$.

\textbf{Theorem 3}. For all $p,q\in \Sigma_3$, $p\equiv_d q \Leftrightarrow p =_d q$.

\begin{proof}
The proof needs to use induction over the depth of derivations. Comparison to the proof of Theorem 2, we only consider the new cases.

($\Leftarrow$).

The new cases are as follows.

1. $\frac{p\approx q}{p=_d q}$. If $p\approx q$, then by use of the last four axioms in Def.12.6 (same as Def.8.6) and the five axioms in Def.12.7, we have $p\equiv_0 q$, by Def.12.1 we have $q\equiv_d q$, finally by Def.12.3, we get $p\equiv_d q$.

2. $\frac{p=_d p'}{p\mid q=_d p'\mid q}$. By the i.e., $p\equiv_d p'$; by Def.12.1, we have $q\equiv_0 q$; then we apply Def.12.7, we get $p\mid q=_d p'\mid q$.

($\Rightarrow$).

1. $p\equiv_d p$. We have $p\approx p$, then we get $p=_d p$.

2. $p\equiv_{d_1} p', q\equiv_{d_2} q' \Rightarrow p\mid q\equiv_{d_1+d_2} p'\mid q'$ for $d_1,d_2<\theta$. By the i.e., we get $p=_{d_1} p'$ and $q=_{d_2} q'$; by use of Def.11.3 and Def.11.5, we get $p\mid q=_{d_1+d_2} p'\mid q'$.

3. If $u$ is $\sum \mu_i x_i$ and $v$ is $\sum \nu_j y_j$, then $u\mid v\equiv_d\sum \mu_i(x_i\mid v) + \sum \nu_j(u\mid y_j) + \sum_{\overline{d}(\mu_i,\nu_j)<\theta} \tau(x_i\mid y_j)$ (where $d=\overline{d}(\mu_i,\nu_j),\overline{d}(\mu_i,\nu_j)<\theta$). By the i.e. and Def.11.1, Def.11.2, Def.11.3, and Def.11.5.

4. The four axioms in Def.12.8, by applying the i.h. and Def.11.1, Def.11.2, Def.11.3, and Def.11.5.
\end{proof}

\section{Distances between Processes with Restriction $\backslash$ and Relabeling $f$}

In this section, we will process $\Sigma_4(\textbf{0},Act\cup\overline{Act}\cup\{\tau\},+,\mid, \backslash,f)$, where $\backslash$ is the restriction and $f$ is the relabeling operator.

\textbf{Definition 13 (Distance relation of $\Sigma_4$-algebra processes)}. We just add the following rules to Definition 11 as rule 6 and rule 7.

6.$\frac{p\sqsubseteq^{\mathcal{L}}_{n}p'}{p\backslash L\sqsubseteq^{\mathcal{L}}_{n}p'\backslash L}$ for $L$ is strongly d-round.

7.$\frac{p\sqsubseteq^{\mathcal{L}}_{n}p'}{p[f]\sqsubseteq^{\mathcal{L}}_{n}p'[f]}$ for $f$ is non-expansive.

\textbf{Definition 14 (Algebraic Definition of the bisimulation distance of $\Sigma_4$-algebra)}. We just add the following rules as 9 and 10 to Definition 12:

9.(i)$p\equiv_d p'\Rightarrow p\backslash L\equiv_d p'\backslash L$, if $L$ is strongly d-round.

(ii)$p\equiv_d p'\Rightarrow p[f]\equiv_d p'[f]$, if $f$ is non-expansive.

10.(i)$(\alpha x)[f]\equiv_0 f(\alpha)(x[f])$, if $f(\alpha)$ is defined; otherwise, $(\alpha x)[f]\equiv_0 \textbf{0}$.

(ii)$(x+y)[f]\equiv_0 x[f] + y[f]$.

(iii)$\textbf{0}[f]\equiv_0 \textbf{0}$.

\textbf{Theorem 4}. For all $p,q\in \Sigma_4$, $p\equiv_d q \Leftrightarrow p =_d q$.

\begin{proof}
The proof needs to use induction over the depth of derivations. Comparison to the proof of Theorem 3, we only consider the new cases.

($\Leftarrow$).

The new cases are as follows.

1. $\frac{p\approx q}{p=_d q}$. If $p\approx q$, then by use of the last four axioms in Def.14.6 (same as Def.8.6), the five axioms in Def.14.7, and the three axioms in Def.14.10, we have $p\equiv_0 q$, by Def.14.1 we have $q\equiv_d q$, finally by Def.14.3, we get $p\equiv_d q$.

2. $\frac{p=_d p'}{p\backslash L=_d p'\backslash L}$. By the i.e., $p\equiv_d p'$; then, we have $p\backslash L\equiv_d p'\backslash L$; then we apply Def.14.9(i), we get $p\backslash L=_d p'\backslash L$.

3. $\frac{p=_d p'}{p[f]=_d p'[f]}$. By the i.e., $p\equiv_d p'$; then, we have $p[f]\equiv_d p'[f]$; then we apply Def.14.9(ii), we get $p[f]=_d p'[f]$.

($\Rightarrow$).

1. $p\equiv_d p$. We have $p\approx p$, then we get $p=_d p$.

2. $p\equiv_d p'\Rightarrow p\backslash L\equiv_d p'\backslash L$, if $L$ is strongly d-round. By the i.e. and Def.13.6.

3. $p\equiv_d P'\Rightarrow p[f]\equiv_d p'[f]$, if $f$ is non-expansive. By the i.e. and Def.13.7.

4. The three axioms in Def.14.10, by applying the i.h. and Def.13.1, Def.13.2, Def.13.3 and Def.11.7.
\end{proof}

\section{Distances between Processes with Recursion}

In this section, we continue discussing $\Sigma_5(\textbf{0},Act\cup\overline{Act}\cup\{\tau\},+,\mid, \backslash,f,\overset{\text{def}}{=})$.

\textbf{Definition 15 (Distance relation of $\Sigma_5$-algebra processes)}. We just add the following rules to Definition 13 as rule 8.

8.$\frac{p\sqsubseteq^{\mathcal{L}}_{n}p'}{A\sqsubseteq^{\mathcal{L}}_{n}p'}$ ($A \overset{\text{def}}{=}p$), if $p,p'$ are weakly guarded, $d<\theta$; for any $\lambda>d$, if $X$ is $\lambda$-weakly defended in $p,p'$.

\textbf{Definition 16 (Algebraic Definition of the bisimulation distance of $\Sigma_5$-algebra)}. We just add the following rules as 11 to Definition 14:

11.$p\equiv_d p'\Rightarrow A\equiv_d p'$ ($A \overset{\text{def}}{=}P$), if $p,p'$ are weakly guarded, $d<\theta$; for any $\lambda>d$, if $X$ is $\lambda$-weakly defended in $p,p'$.

\textbf{Theorem 5}. For all $p,q\in \Sigma_5$, $p\equiv_d q \Leftrightarrow p =_d q$.

\begin{proof}
The proof needs to use induction over the depth of derivations. Comparison to the proof of Theorem 4, we only consider the new cases.

($\Leftarrow$).

The new cases are as follows.

1. $\frac{p\approx q}{p=_d q}$. If $p\approx q$, then by use of the last four axioms in Def.16.6 (same as Def.8.6), the five axioms in Def.16.7, and the three axioms in Def.16.10, we have $p\equiv_0 q$, by Def.16.1 we have $q\equiv_d q$, finally by Def.16.3, we get $p\equiv_d q$.

2. $\frac{p=_d p'}{A=_d p'}$. By the i.e., $p\equiv_d p'$; then, we have $A\equiv_d p'$; then we apply Def.16.11, we get $A=_d p'$.

($\Rightarrow$).

1. $p\equiv_d p$. We have $p\approx p$, then we get $p=_d p$.

2. $p\equiv_d p'\Rightarrow A\equiv_d p'$, ($A \overset{\text{def}}{=}p$), if $p,p'$ are weakly guarded, $d<\theta$; for any $\lambda>d$, if $X$ is $\lambda$-weakly defended in $p,p'$. By the i.e. and Def.15.8.
\end{proof}

\section{Conclusions}\label{Conclusions}

Based on the work for distances between basic processes in \cite{DPA}, and Ying's work on approximate bisimilary \cite{AB} \cite{TOPO}, we develop an algebraic approach for a whole process calculus CCS \cite{CCS}.

In future, big data will become more and more popular, and we will try to use the algebraic approach in this paper, to verify the correctness of behaviors of big data systems and applications.

\label{lastpage}


\begin{thebibliography}{Lam94}

  \bibitem{CCS} R. Milner.: \emph{Communication and Concurrency.}
    Prentice Hall, 1989.

  \bibitem{CCS2} R. Milner and J. Parrow and D. Walker.: \emph{A calculus of mobile processes, Parts I and II.}
    Information and Computation, 1992, 100(1992): 1--77.

  \bibitem{HML} M. Hennessy, R. Milner.: \emph{Algebraic Laws for Nondeterminism and Concurrency.} J. ACM, 1985, 32(1): 137--161.

  \bibitem{AB} M. Ying, M. Wirsing.: \emph{Approximate bisimilarity.} In: Rus, T. (ed.) AMAST 2000. LNCS, vol. 1816, pp. 309¨C322. Springer, Heidelberg (2000).

  \bibitem{TOPO} M. Ying.: \emph{Topology in process calculus - approximate correctness and infinite evolution of concurrent programs.} Springer (2001).

  \bibitem{SD}P. \v{C}ern\'{y}, T. A. Henzinger, A. Radhakrishna.: \emph{Simulation distances.} Theoretical Computer Science 413.1(2012):21¨C35.

  \bibitem{DPS} D. R. Hern\'{a}ndez, D. D. F. Escrig.: \emph{Defining distances for all process semantics.} Proceedings of the 14th joint IFIP WG 6.1 international conference and Proceedings of the 32nd IFIP WG 6.1 international conference on Formal Techniques for Distributed Systems Springer-Verlag, 2012:169-185.

  \bibitem{DPA} D. R. Hern\'{a}ndez, D. D. F. Escrig.: \emph{Distances between Processes: A Pure Algebraic Approach.} Recent Trends in Algebraic Development Techniques. Springer, Berlin Heidelberg, 2013:265-282.

\end{thebibliography}
\end{document}